# Entropic Cohesion in Vitrimers

Rahul Karmakar,[1] Himanshu,[1] Srikanth Sastry,[2*] Sanat K. Kumar,[3*] and Tarak K. Patra[1*]

[1]Department of Chemical Engineering, Indian Institute of Technology, Madras, Chennai 600036, India
[2]Theoretical Sciences Unit and School of Advanced Materials, Jawaharlal Nehru Centre for Advanced Scientific Research, Rachenahalli Lake Road, Bengaluru-560064, India
[3]Department of Chemical Engineering, Columbia University, New York, USA

## Abstract

Vitrimers – polymer networks that can undergo bond exchange reactions – dynamically rearrange their structures while maintaining their overall integrity, thus resulting in unique properties such as self-healing, reprocessability, shape memory and adaptability. Here, we show that the introduction of dynamic bonds directly impacts the polymer's density. For a limiting case, where the dynamic bonds are the same size as the polymer chain bonds, simulations and theory show an enhancement in the density, because these bonds induce an increased cohesive force in the liquid, which is entropic in origin. The crosslinks are well mixed in the bulk but are depleted from the air-polymer interface. These findings implicate density as a key variable in polymers with dynamic crosslinkers, one that can be used to facilely tune their properties.

*Authors to correspond: TKP: tpatra@iitm.ac.in
SKK: sk2794@columbia.edu
SS: sastry@jncasr.ac.in

Permanently crosslinked polymers are extensively used due to their favorable mechanical properties. However, these materials are unsustainable since they cannot be reprocessed. In contrast, polymer networks that can rearrange their topology without depolymerization, e.g., via the formation of reversible crosslinks ("vitrimers"), are reprocessable, and hence are promising candidates for a circular economy.[1–13] The dynamic bond exchange in vitrimers is facilitated by reversible chemical reactions, such as transesterification or disulfide bond exchange, which enables favorable mechanical, rheological, self-healing, adhesive, and shape memory properties.[14,15] In a recent study, we found that reversible cross-links can change the miscibility of polymer blends.[16] Since blend miscibility can be intimately affected by changes in polymer density, we ask as to how the introduction of dynamic crosslinks affects the equilibrium polymer density (which is in coexistence with its vapor).

To critically address this question here we study the structure and dynamics of a coarse grained model polymer melt with exchangeable crosslinks using a hybrid molecular simulation method. The bond exchange is simulated using a Monte Carlo (MC) scheme, while molecular dynamics (MD) simulations are conducted to relax the system. We find that the saturated liquid density of the polymer increases monotonically with the fraction of crosslinked monomers. Also, we find that crosslinks are preferentially segregated away from the free surface. Since both of these results are not energetic in their origin, we propose that crosslinks provide entropic cohesion to the vitrimer. Since many properties of a vitrimer (e.g., miscibility, fragility) are critically determined by its density, we propose that varying the crosslinking fraction is a facile means of tuning their practically relevant behavior.

We use the Kremer-Grest bead-spring polymer model[17] and perform hybrid MD – MC simulations in the isothermal-isobaric ensemble. The number of monomers in a chain is varied systematically in the range $N=5$ to 100 in a series of simulations; most of our computations focus on $N=20$. Catenated monomers on a chain are held together by a standard finite extensible nonlinear elastic (FENE) bond.[17] The total number of monomers is 10000. The interaction between a pair of monomers is described by the Lennard-Jones (LJ) potential truncated at $2.5\sigma$ ($\sigma$ is the monomer diameter). The simulation box is periodic in all three directions, but the 3rd direction ($z$-axis) is significantly longer that the other two so that we have a slab of liquid exposed to vacuum along this direction (Figure 1a).[18] The formation/breaking of crosslink bonds are modelled using a configurational-biased MC algorithm as has recently been proposed by Qin and his coworkers.[19] During the MD we perform 200 MC moves (defined as a MC cycle) at regular intervals - the MD time between two consecutive MC cycles is denoted as $\tau_c$. During each MC step, either two randomly chosen monomers within a cutoff distance ($1.5\sigma$) are connected by the FENE potential as discussed above, or a randomly chosen existing crosslink is deleted. (Main chain bonds are never created or destroyed.) The bond creation/destruction is accepted following the Metropolis algorithm, where $\mu$, the chemical potential determines the equilibrium fraction of crosslinks. The monomers only move during the MD, and are not allowed to move during this bond formation/destruction step. We first perform $10^7$ MD steps with an integration timestep of $0.005\tau$ to equilibrate the system ($t = 5 \times 10^4 \tau$), followed by a production cycle of $10^7$ MD steps (see below). Here, $\tau = \sqrt{m\sigma^2/\epsilon}$ is the unit of time, $m$, $\sigma$ and $\epsilon$ are the mass, size and interaction energy of a monomer, respectively. The temperature of the system $\left(\frac{k_B T}{\epsilon}\right) = 1$ is maintained by the Nose-Hover thermostat. MD simulations are conducted using the LAMMPS open-source code.[20]

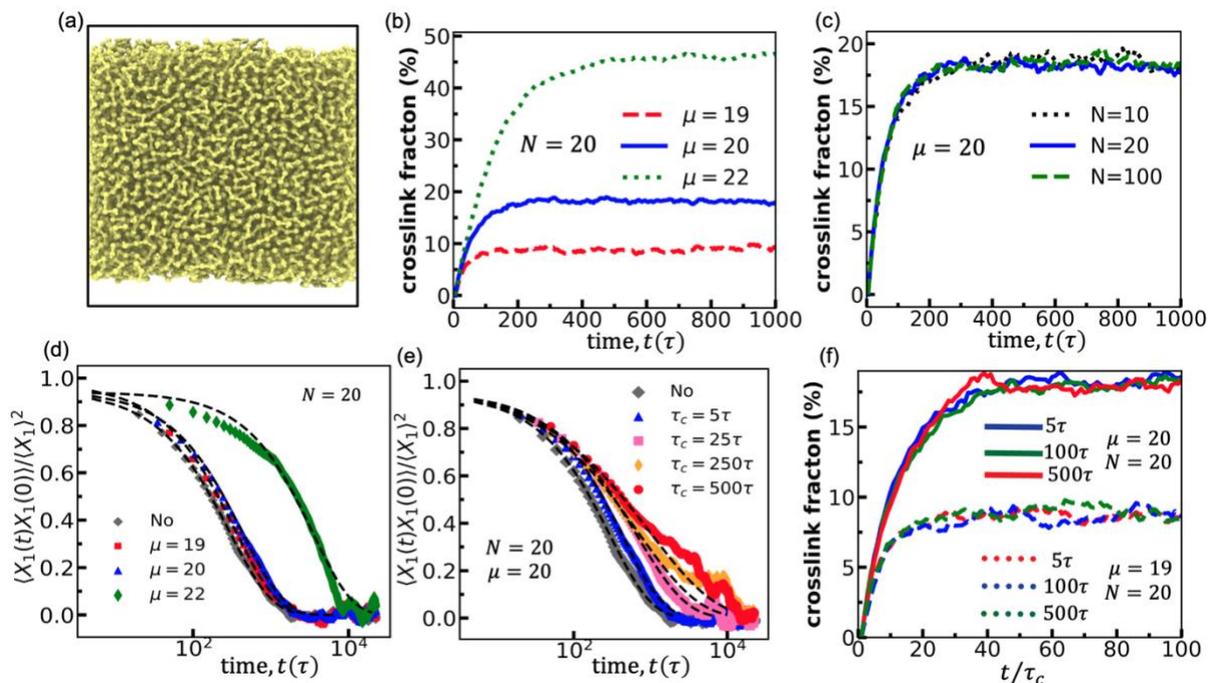

*Figure 1: Modeling reversible dynamic crosslinks in a polymer melt. An MD snapshot of the simulation box is shown in (a). The fraction of particles that are dynamically crosslinked during the MD is shown in (b) for N=20 for three different chemical potentials, µ. For a given µ, the crosslink fraction vs. MD time is shown for three different chain lengths in (c). The end-to-end vector autocorrelation functions are plotted in (d) and (e) for different µ and $\tau_c$, respectively. The crosslink fractions are plotted as a function of MD time normalized by $\tau_c$ in (f).*

Our simulation protocol successfully generates a polymer network with an equilibrium crosslink density for each $\mu$ examined (Figure 1b). This crosslink fraction is found to be effectively independent of $N$ (Figure 1c) at a fixed $\mu$, but increases with increasing $\mu$. The time dependent chain end-to-end distance correlation function (Figure 1d and e) clearly shows that all our simulations are long enough to relax the chains and achieve equilibration.[21] Evidently, the chain relaxation time increases as a function of $\mu$. The time evolution of the crosslink fraction is also influenced by $\tau_c$, as shown in Figure 1f. Since these results collapse onto a single curve as a function of $t/\tau_c$ for each $\mu$, apparently, chain relaxation phenomena do not set the time scale for crosslink equilibration.[13]

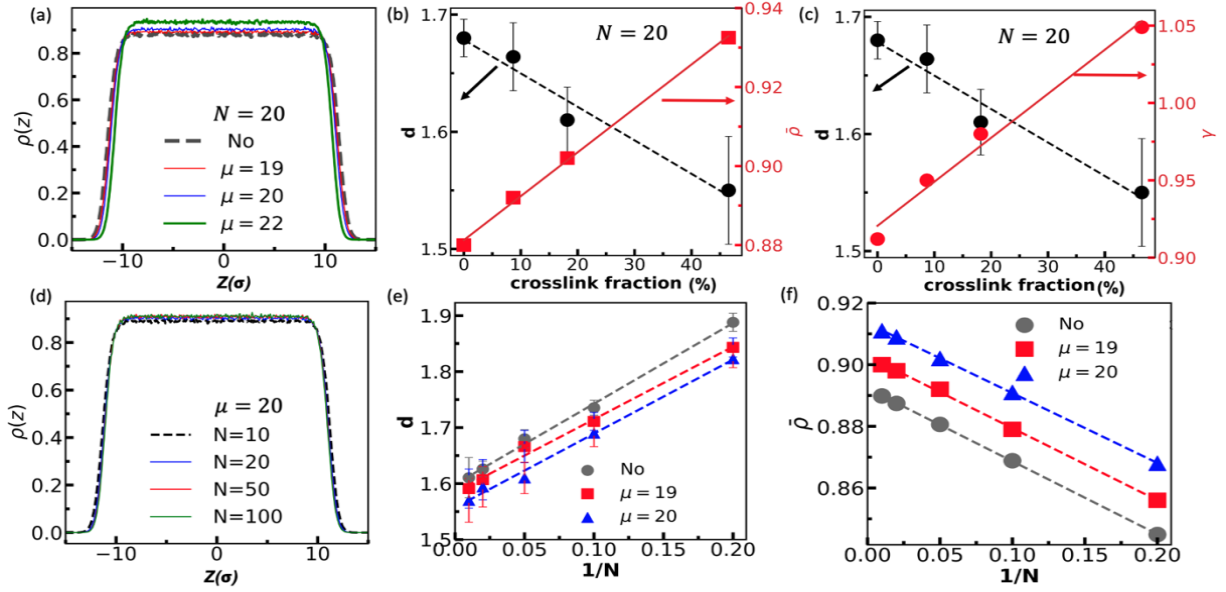

*Figure 2: Thermophysical properties of the dynamically crosslinked polymer. (a) The monomer density profile for N=20. The interfacial width and bulk density is plotted as a function of crosslink fraction in (b). The interfacial width and surface tension is plotted as a function of crosslink fraction in (c). The monomer density profile for three different N for a given μ is shown in (d). The interfacial width and density are plotted as a function of $1/N$ in (e) and (f), respectively.*

We now focus on the structure and interfacial properties of the vitrimer. The monomer density profile along the film thickness direction, $z$, is shown in Figure 2a for $N=20$. It is clear that adding crosslinks increases the bulk polymer density (Figure 2b). This is a key result of our work. We calculate the interfacial width of the film by fitting a hyperbolic tangent curve to the density profile data and report the results in Figure 2b. With increasing crosslink fraction, the interfacial width decreases consistent with the density increases. We estimate the interfacial tension as $\gamma = \frac{L_z}{2}\left[p_{zz} - \frac{p_{xx}+p_{yy}}{2}\right]$, wherein $p_{xx}$, $p_{yy}$ and $p_{zz}$ are the pressure components along $x$, $y$, $z$ directions of the simulation box, respectively.[22] Here, the $L_z$ is the box length along the $z$ direction. We calculate the components of the pressure tensor as $p_{ij} = \frac{1}{V}\sum_{k=1}^{M} m_k v_{ik} v_{jk} + \frac{1}{V}\sum_{k=1}^{M} r_{ik} f_{jk}$ wherein $v$, $r$ and $f$ denote the velocity, position and force components of a monomer, respectively. The indices $i$ and $j$ refer to axes of the simulation box ($x$, $y$ and $z$), and the index $k$ runs over all the monomers in the simulation box ($M$). The volume of simulation box is $V$. The interfacial tension is also found to increase linearly with crosslink fraction, Figure 2c. The interfacial width, tension and bulk density vary linearly with $1/N$ (Figures 2e and f) as expected. It is evident that increasing the crosslink fraction causes the bulk density vs. $1/N$ curves to be shifted up. We thus conjecture that the crosslinks increase the cohesion of the

polymers – since this is associated with bond addition, but no significant additional changes in the system energetics, we postulate that this increased cohesion is entropic in origin.

To bolster these arguments, we have performed calculations on a one-dimensional hard rod system (see Supplementary information), which clearly shows that the addition of bonding constraints serves to increase its density.[23,24] This gives considerable confidence in our conjecture that this extra cohesion in the vitrimer case is also entropic in its origin. In contrast to these definitive theoretical statements, the experimental literature on role of crosslinks on polymer density is sketchy and contradictory.[25–27] There is essentially no literature on the role of dynamic crosslinks on the saturated polymer density. In contrast, polymer density has been found to both decrease and increase for permanently crosslinked systems, apparently determined by the chemical nature of the crosslinkers in question. These results are further complicated by the fact that such permanently crosslinked systems, especially those above their percolation thresholds, are frequently out of equilibrium. Thus, at this time we do not have experimental validation for our ideas. Regardless, we conjecture that the bulkiness of the crosslinkers could be played off against the entropic cohesion offered by dynamic crosslinking. This is a particularly interesting direction since previous theoretical work has suggested that small variations in vitrimer density can cause large changes to system fragility. Additionally, since the cohesive energy density of a polymer is strongly affected by density,[28] we argue that variations of crosslinker bulkiness can be used to systematically control the miscibility of polymer blends to which they are added.

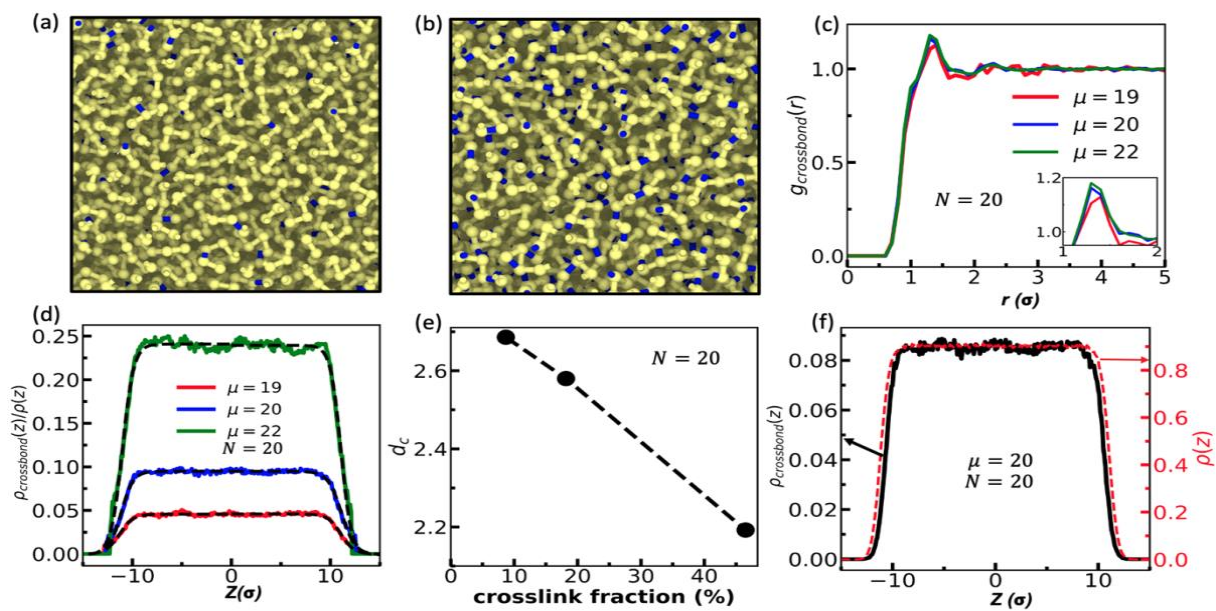

*Figure 3: MD snapshots of dynamically crosslinked polymer films are shown in (a) and (b) for crosslink fraction of 20% and 45%, respectively. The crosslinks are in blue. The crosslink bond correlation function is shown in (c). The inset of (c) represents a close-up view of the first peak. The normalized crosslink density across the film is plotted in (d). The dotted lines correspond to tanh fits. The resulting interfacial width is plotted in (e) as a function of crosslink fraction. The crosslinker density and monomer density profiles are compared in (f).*

Figures 3a-b shows MD snapshots of the system with two different chemical potentials, wherein the crosslink bonds are highlighted in blue. The bond-bond pair correlation function in the bulk region of the film (Figure 3c) suggests an essentially uniform distribution of crosslinks. The excluded volume of the crosslinks[29] $V_C = -\int_0^\infty [g(r) - 1] 4\pi r^2 dr > 0$ for all the cases studied, indicating that the crosslinks repel each other. This finding clearly suggests that the recent results of Richert et al.[30], who found that dynamic crosslinks in a polyethylene melt segregated into domains, cannot be reproduced by the purely entropic arguments presented here. Rather, energetic effects, steric or polymer crystallization must be additional drivers not accounted for in our calculations.

Finally, to provide another piece of evidence in favor of our entropic cohesion argument, we examine the crosslink density normalized by the particle density across the thin film (Figure 3d). The resulting normalized crosslink density profiles also follow a hyperbolic tangent form, but the interfacial width of this quantity is significantly larger than the values shown for the density profiles in Figure 2. This increased width, which also varies inversely with the crosslink fraction (Figure 3e), illustrates that the crosslinks are preferentially depleted at the gas-liquid interface. This depletion of crosslinks is further illustrated in Figure 3f.

In summary, vitrimers have the ability to undergo reversible bond exchange, making them highly adaptable and repairable. Here, we report that this reversible crosslinking mechanism can be used to increase polymer cohesion – this increases the density and makes the gas-liquid interface sharper. These findings indicate that density is a key controllable variable in polymers with dynamic crosslinkers, one that can be used to facilely tune their properties.

**Acknowledgements**

This work is made possible by financial support from the SERB, DST, and Govt of India through a core research grant (CRG/2022/006926) and the National Supercomputing Mission's research grant (DST/NSM/R&D_HPC_Applications/2021/40). This research uses the resources of the Center for Nanoscience Materials, Argonne National Laboratory, which is a DOE Office of Science User Facility supported under the Contract DE-AC02-06CH11357. We thank Jian Qin and Douglas Li for the help in implementing the MC-MD simulation of vitrimers.

**References:**

1. Zhao, H. *et al.* Molecular Dynamics Simulation of the Structural, Mechanical, and Reprocessing Properties of Vitrimers Based on a Dynamic Covalent Polymer Network. *Macromolecules* (2022) doi:10.1021/acs.macromol.1c02034.

2. Zhao, H. *et al.* Unveiling the Multiscale Dynamics of Polymer Vitrimers Via Molecular Dynamics Simulations. *Macromolecules* **56**, 9336–9349 (2023).

3. Montarnal, D., Capelot, M., Tournilhac, F. & Leibler, L. Silica-Like Malleable Materials from Permanent Organic Networks. *Science* **334**, 965–968 (2011).

4. Lu, Y.-X., Tournilhac, F., Leibler, L. & Guan, Z. Making Insoluble Polymer Networks Malleable via Olefin Metathesis. *J. Am. Chem. Soc.* **134**, 8424–8427 (2012).

5. Snyder, R. L., Fortman, D. J., De Hoe, G. X., Hillmyer, M. A. & Dichtel, W. R. Reprocessable Acid-Degradable Polycarbonate Vitrimers. *Macromolecules* **51**, 389–397 (2018).

6. Jin, Y., Yu, C., Denman, R. J. & Zhang, W. Recent advances in dynamic covalent chemistry. *Chem. Soc. Rev.* **42**, 6634–6654 (2013).

7. Samanta, S., Kim, S., Saito, T. & Sokolov, A. P. Polymers with Dynamic Bonds: Adaptive Functional Materials for a Sustainable Future. *J. Phys. Chem. B* **125**, 9389–9401 (2021).

8. Maaz, M., Riba-Bremerch, A., Guibert, C., Van Zee, N. J. & Nicolaÿ, R. Synthesis of Polyethylene Vitrimers in a Single Step: Consequences of Graft Structure, Reactive Extrusion Conditions, and Processing Aids. *Macromolecules* **54**, 2213–2225 (2021).

9. Long, R., Qi, H. J. & Dunn, M. L. Modeling the mechanics of covalently adaptable polymer networks with temperature-dependent bond exchange reactions. *Soft Matter* **9**, 4083–4096 (2013).

10. Meng, F., Pritchard, R. H. & Terentjev, E. M. Stress Relaxation, Dynamics, and Plasticity of Transient Polymer Networks. *Macromolecules* **49**, 2843–2852 (2016).

11. Ricarte, R. G. & Shanbhag, S. Unentangled Vitrimer Melts: Interplay between Chain Relaxation and Cross-link Exchange Controls Linear Rheology. *Macromolecules* **54**, 3304–3320 (2021).

12. Semenov, A. N. & Rubinstein, M. Thermoreversible Gelation in Solutions of Associative Polymers. 1. Statics. *Macromolecules* **31**, 1373–1385 (1998).

13. Rubinstein, M. & Semenov, A. N. Thermoreversible Gelation in Solutions of Associating Polymers. 2. Linear Dynamics. *Macromolecules* **31**, 1386–1397 (1998).

14. Porath, L., Soman, B., Jing, B. B. & Evans, C. M. Vitrimers: Using Dynamic Associative Bonds to Control Viscoelasticity, Assembly, and Functionality in Polymer Networks. *ACS Macro Lett.* 475–483 (2022) doi:10.1021/acsmacrolett.2c00038.

15. Zou, Z. *et al.* Rehealable, fully recyclable, and malleable electronic skin enabled by dynamic covalent thermoset nanocomposite. *Sci. Adv.* **4**, eaaq0508.

16. Clarke, R. W. *et al.* Dynamic crosslinking compatibilizes immiscible mixed plastics. *Nature* **616**, 731–739 (2023).

17. Kremer, K. & Grest, G. S. Molecular dynamics (MD) simulations for polymers. *J. Phys. Condens. Matter* **2**, SA295–SA298 (1990).

18. Madden, W. G. Monte Carlo studies of the melt–vacuum interface of a lattice polymer. *J. Chem. Phys.* **87**, 1405–1422 (1987).

19. Li, D. T., Rudnicki, P. E. & Qin, J. Distribution Cutoff for Clusters near the Gel Point. *ACS Polym. Au* (2022) doi:10.1021/acspolymersau.2c00020.

20. Thompson, A. P. *et al.* LAMMPS - a flexible simulation tool for particle-based materials modeling at the atomic, meso, and continuum scales. *Comput. Phys. Commun.* **271**, 108171 (2022).

21. Kalathi, J. T., Kumar, S. K., Rubinstein, M. & Grest, G. S. Rouse mode analysis of chain relaxation in polymer nanocomposites. *Soft Matter* **11**, 4123–4132 (2015).

22. Meenakshisundaram, V., Hung, J.-H., Patra, T. K. & Simmons, D. S. Designing Sequence-Specific Copolymer Compatibilizers Using a Molecular-Dynamics-Simulation-Based Genetic Algorithm. *Macromolecules* **50**, 1155–1166 (2017).

23. Bryngelson, J. D. & Thirumalai, D. Internal Constraints Induce Localization in an Isolated Polymer Molecule. *Phys. Rev. Lett.* **76**, 542–545 (1996).

24. Corti, D. S. & Debenedetti, P. G. Statistical mechanics of fluids under internal constraints: Rigorous results for the one-dimensional hard rod fluid. *Phys. Rev. E* **57**, 4211–4226 (1998).

25. Fox, T. G. & Loshaek, S. Influence of molecular weight and degree of crosslinking on the specific volume and glass temperature of polymers. *J. Polym. Sci.* **15**, 371–390 (1955).

26. Charlesby, A., Ross, M. & Freeth, F. A. The effect of cross-linking on the density and melting of Polythene. *Proc. R. Soc. Lond. Ser. Math. Phys. Sci.* **217**, 122–135 (1997).

27. Lenhart, J. L. & Wu, W. Influence of Cross-Link Density on the Thermal Properties of Thin Polymer Network Films. *Langmuir* **19**, 4863–4865 (2003).

28. White, R. P. & Lipson, J. E. G. Free Volume, Cohesive Energy Density, and Internal Pressure as Predictors of Polymer Miscibility. *Macromolecules* **47**, 3959–3968 (2014).

29. Hansen, J.-P. & McDonald, I. R. *Theory of Simple Liquids*. (Academic Press, 2006).

30. Ricarte, R. G., Tournilhac, F. & Leibler, L. Phase Separation and Self-Assembly in Vitrimers: Hierarchical Morphology of Molten and Semicrystalline Polyethylene/Dioxaborolane Maleimide Systems. *Macromolecules* **52**, 432–443 (2019).